\documentclass[10pt,conference]{IEEEtran}
\IEEEoverridecommandlockouts

\usepackage{cite}
\usepackage[utf8]{inputenc}
\usepackage{amsmath,amssymb,amsfonts}
\usepackage{algorithm}
\usepackage{algorithmicx}
\usepackage{algpseudocode}
\usepackage{graphicx}
\usepackage{textcomp}
\usepackage{xcolor}
\usepackage{balance}
\usepackage{makecell}
\usepackage{pifont}
\usepackage{url}
\usepackage{booktabs}
\usepackage{tabularx}
\usepackage{enumitem}
\usepackage{hyperref} 
\usepackage{subcaption}
\DeclareRobustCommand{\myurl}[1]{\url{#1}}

\def\BibTeX{{\rm B\kern-.05em{\sc i\kern-.025em b}\kern-.08em T\kern-.1667em\lower.7ex\hbox{E}\kern-.125emX}}

\newcommand{\todo}[1]{\textsf{\textbf{\textcolor{Orange}{[#1]}}}}

\newcommand{\boldpar}[1]{\smallbreak\noindent\textbf{#1.}}

\begin{document}
\title{
Latency-aware Human-in-the-Loop Reinforcement Learning for Semantic Communications
}

\author{\IEEEauthorblockN{
Peizheng Li, Xinyi Lin, Adnan Aijaz
}\\ 
\IEEEauthorblockA{
Bristol Research and Innovation Laboratory, Toshiba Europe Ltd., U.K.\\
Email: 
{\{peizheng.li, xinyi.lin, adnan.aijaz\}@toshiba.eu}}
}

\maketitle

\begin{abstract}
Semantic communication promises task-aligned transmission but must reconcile semantic fidelity with stringent latency guarantees in immersive and safety-critical services. This paper introduces a time-constrained human-in-the-loop reinforcement learning (TC-HITL-RL) framework that embeds human feedback, semantic utility, and latency control within a semantic-aware Open radio access network (RAN) architecture. We formulate semantic adaptation driven by human feedback as a constrained Markov decision process (CMDP) whose state captures semantic quality, human preferences, queue slack, and channel dynamics, and solve it via a primal--dual proximal policy optimization algorithm with action shielding and latency-aware reward shaping. The resulting policy preserves PPO-level semantic rewards while tightening the variability of both air-interface and near-real-time RAN intelligent controller processing budgets. Simulations over point-to-multipoint links with heterogeneous deadlines show that TC-HITL-RL consistently meets per-user timing constraints, outperforms baseline schedulers in reward, and stabilizes resource consumption, providing a practical blueprint for latency-aware semantic adaptation.
\end{abstract}


\section{Introduction}
Semantic communication (SemCom) shifts the design focus from bit-level fidelity to task- or meaning-level utility, transmitting only task-relevant information and enabling joint design of physical, link, and inference layers for improved spectral and energy efficiency as well as reduced latency~\cite{9663101,10597087,lin2025rl}. Specifically, deep learning–based SemCom systems, often realized via joint source–channel coding (JSCC)~\cite{bourtsoulatze2019deep,gunduz2024joint}, have demonstrated strong robustness to channel impairments and performance gains. However, most existing designs treat semantic models as \emph{static} once trained and therefore struggle to maintain alignment when wireless conditions, user preferences, or task objectives evolve over time. From a service perspective, adaptive mechanisms are essential to keep semantic fidelity aligned with user intent and application context.

The rapid progress of generative AI and reinforcement learning from human feedback (RLHF)~\cite{lambert2025reinforcement} underscores the value of learning directly from human preferences. Human-in-the-Loop Reinforcement Learning (HITL-RL) incorporates subjective feedback into the reward design and policy updates \cite{retzlaff2024human}. It has been successfully applied in robotics, preference learning, and controllable text generation, and has recently been advocated for SemCom to align models with user-perceived utility~\cite{10849728,11152855}. Bringing HITL-RL into networked SemCom loops, however, introduces domain-specific challenges.

In wireless systems, human feedback travels over bandwidth- and latency‑limited links, and semantic model updates must meet strict timing constraints. In point‑to‑multipoint deployments with heterogeneous users, feedback delays and reconfiguration latencies can render otherwise beneficial updates \emph{infeasible} for a subset of users. Ignoring these temporal effects leads to per‑user deadline violations and degraded quality of experience (QoE). Time‑aware decision mechanisms are therefore required to couple semantic utility with the realities of scheduling and deployment. Meanwhile, the granularity of model updates (e.g., partial refresh vs. full retraining) should be carefully chosen to balance semantic gains against latency overhead.

Constrained Markov decision processes (CMDPs)~\cite{altman2021constrained} provide a principled way to enforce latency or safety budgets via Lagrangian or primal–dual methods~\cite{achiam2017constrained}. Proximal Policy Optimization (PPO) \cite{schulman2017proximal}, known for stability and sample efficiency, can be endowed with cost critics and dual variables to form constrained PPO (PPO-C), and recent work has brought such RL ideas to RIC optimization~\cite{9931127,li2025toward}. However, prior studies either focus on average QoS or resource slicing, without incorporating human preference signals or per-frame feasibility mechanisms as introduced here.

In this work, we introduce a \emph{time-constrained} HITL-RL framework for semantic adaptation in point-to-multipoint settings. We formulate semantic adaptation as a CMDP with per-user deadline budgets and latency-aware reward shaping, and we solve it with a primal–dual PPO algorithm augmented with an \emph{action shield} that enforces instantaneous feasibility during both training and deployment. To our knowledge, this is among the first integrations of HITL-RL with SemCom under explicit real-time constraints, bridging preference-driven learning with implementable timing control.
The main contributions are:
\begin{itemize}
    \item \textbf{Latency-aware CMDP.} We couple human-aligned semantic utility with near-RT RIC budgets and per-user deadlines, yielding a tractable CMDP abstraction for semantic broadcasting under latency guarantees.
    \item \textbf{TC-PPO with shielding.} A primal--dual PPO variant with cost critics, adaptive multipliers, and an action shield enforces both average and instantaneous feasibility.
    \item \textbf{Implementation and evidence.} We map the framework to an NR-like slot structure and show on JSCC-enabled multi-user links that TC-PPO preserves PPO-level reward while tightening resource dispersion; targeted ablations highlight the role of each component.
\end{itemize}

\section{System Model}
\label{sec: system model}

\begin{figure}[t]
    \centering
    \includegraphics[width=0.9\linewidth]{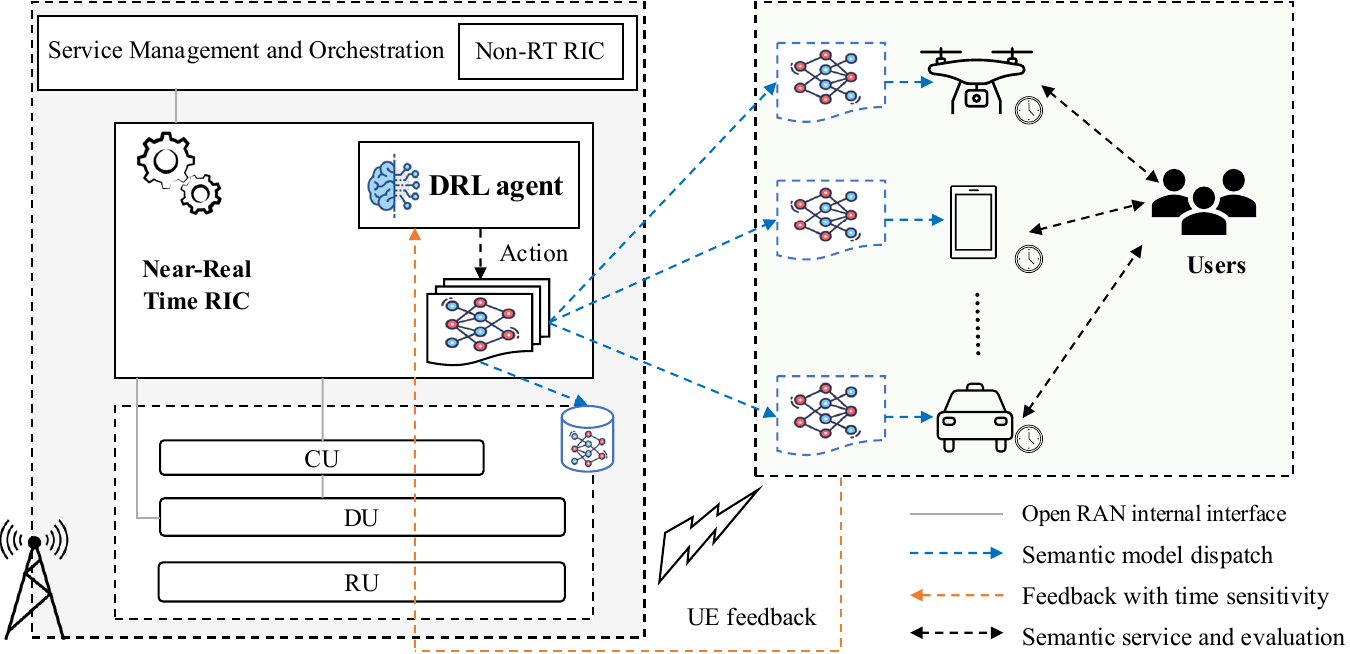}%
    \caption{Illustration of the system model.}
    \label{fig:system_model}
\end{figure}
%

We consider an AI-driven next-generation RAN where a semantic-aware base station (gNB) serves latency-heterogeneous user equipments (UEs) $\mathcal{N}=\{1,\ldots,N\}$ over a shared downlink. As illustrated in Fig.~\ref{fig:system_model}, the architecture follows the Open RAN functional split~\cite{10597087,li2023open}: the near-RT RIC hosts the HITL-RL agent, while distributed units (DUs) and radio units (RUs) handle physical-layer connectivity. Semantic models operate as encoder–decoder pairs, with the encoder at the gNB and UE-specific decoders at the terminals. Human operators evaluate reconstructed semantics and send preference feedback to the RIC, which fuses these signals, updates the models, and disseminates configuration deltas under strict timing budgets.

The control loop is discretized into gNB scheduling sub-frames indexed by $t \in \{0,1,\ldots\}$, and each sub-frame comprises slot grants (akin to NR mini-slot allocations) that are dynamically carved out for semantic adaptation. Every UE $i$ belongs to a service class $k(i)$ with a deadline $d_i$ representing the maximum allowable time between observing semantic degradation and deploying a refreshed decoder.

\subsection{Semantic Delivery Pipeline}
At frame $t$ the gNB ingests source features $\mathbf{x}_t \in \mathbb{R}^{n_s}$ together with historical context $\mathbf{m}_t$. The encoder parameters $\phi_t$ generate a latent embedding
\begin{equation}
\mathbf{z}_t = f_{\phi_t}(\mathbf{x}_t, \mathbf{m}_t),
\end{equation}
which is mapped onto a complex-valued symbol block $\mathbf{s}_t = \mathcal{E}(\mathbf{z}_t) \in \mathbb{C}^{n_c}$ satisfying $\|\mathbf{s}_t\|_2^2 \le n_c P_{\max}$ for transmission, where $P_{\max}$ denotes the per-symbol power budget. The RU simultaneously delivers $\mathbf{s}_t$ over a block-fading MIMO channel,
\begin{equation}
\mathbf{y}_{i,t} = \mathbf{H}_{i,t} \mathbf{s}_t + \mathbf{n}_{i,t},
\end{equation}
where $\mathbf{H}_{i,t}$ captures path loss and small-scale fading toward UE $i$ and $\mathbf{n}_{i,t} \sim \mathcal{CN}(\mathbf{0},\sigma_i^2 \mathbf{I})$. Each UE maintains a personalized decoder $\psi_{i,t}$ that incorporates local side information $\mathbf{c}_{i,t}$ (e.g., service context information or sensor snapshots),
\begin{equation}
\hat{\mathbf{x}}_{i,t} = g_{\psi_{i,t}}(\mathbf{y}_{i,t}, \mathbf{c}_{i,t}).
\end{equation}
The reconstruction is assessed through a task loss $\ell_i(\hat{\mathbf{x}}_{i,t}, \mathbf{x}_t)$ and its complementary quality score $q_{i,t} = 1-\ell_i(\hat{\mathbf{x}}_{i,t}, \mathbf{x}_t)$, which is reported to the near-RT RIC.

\subsection{Human Feedback Acquisition}
UE-side human evaluators send scalar or vector feedback $F_{i,t}$ through uplink control or data bearers configured by the gNB. The specific PHY channels used for this purpose depend on deployment choices and are therefore abstracted in this study. Our focus is on how the near-RT RIC acquires, aggregates, and exploits the feedback received at the gNB. Within the Open RAN framework, the SMO and near-RT RIC modules fuse these signals and map them into normalized preference scores,
\begin{equation}
\tilde{U}_{i,t} = (1-\eta_i)\,\mathcal{U}_i(q_{i,t}, \mathbf{p}_{i,t}) + \eta_i \,\mathcal{H}_{\mathrm{pref}}(F_{i,t}),
\end{equation}
where $\mathcal{U}_i$ encodes objective task KPIs $\mathbf{p}_{i,t}$ (e.g., detection accuracy) and $\mathcal{H}_{\mathrm{pref}}$ captures subjective satisfaction, while $\eta_i \in [0,1]$ balances machine and human inputs. The near-RT RIC maintains an exponentially weighted moving average,
$\bar{U}_{i,t+1} = (1-\alpha_i)\bar{U}_{i,t} + \alpha_i \tilde{U}_{i,t},$
with smoothing factor $\alpha_i \in (0,1]$, which becomes part of the RL state and anchors long-term semantic alignment.

\subsection{Latency Budget Decomposition}
Executing an adaptation action $a_t \in \mathcal{U}$, ranging from lightweight decoder-statistic refreshes, through mid-tier feature refinement, to full retraining, cached rollback, or a no-operation (NoOp), triggers an operation sequence in the Open RAN stack. For UE $i$, the end-to-end latency is modeled as
\begin{equation}
\label{eq:c_i_t}
C_{i,t}(u_t) = C^{\mathrm{fb}}_{i,t} + C^{\mathrm{RIC}}_{i,t}(u_t) + C^{\mathrm{tx}}_{i,t}(u_t) + C^{\mathrm{reconf}}_{i,t}(u_t),
\end{equation}
where $C^{\mathrm{fb}}_{i,t}$ denotes feedback acquisition (human response and uplink), $C^{\mathrm{RIC}}_{i,t}$ covers analytics and decision making at the near-RT RIC (including queuing), $C^{\mathrm{tx}}_{i,t}$ accounts for fronthaul/backhaul dissemination of the new model parameters, and $C^{\mathrm{reconf}}_{i,t}$ describes UE-side deployment and warm-start. Hard timing constraints require
\begin{equation}
C_{i,t}(u_t) \le d_i, \quad \forall i \in \mathcal{N}.
\end{equation}
Service classes $k \in \mathcal{K}$ organize UEs by application, yielding representative budgets aligned with 3GPP delay targets in the standardized 5G QoS Identifier (5QI):
\begin{equation}
d_i = T_{k(i)}^{\mathrm{fb}} + T_{k(i)}^{\mathrm{RIC}} + T_{k(i)}^{\mathrm{tx}} + T_{k(i)}^{\mathrm{reconf}}.
\end{equation}
The RIC exposes the residual slack $\Delta_{i,t} = d_i - C_{i,t}(u_t)$ and the normalized deadline debt $\delta_{i,t} = [ -\Delta_{i,t} ]^{+} / d_i$, where $[x]^+ \triangleq \max\{x,0\}$, to the learning agent for scheduling \footnote{Intuitively, $\Delta_{i,t}\!\ge\!0$ indicates remaining time budget before the deadline $d_i$, whereas $\Delta_{i,t}\!<\!0$ quantifies lateness. The normalized debt $\delta_{i,t} \in [0,1]$ expresses any lateness as a fraction of $d_i$.}. 


\subsection{Time Resource Coupling}
Within each 10\,ms gNB frame we assume a fixed numerology $\mu \in \{0,1,2\}$, which sets the slot duration $T_{\mathrm{slot}}(\mu) = 1\,\text{ms}/2^{\mu}$ and symbol duration $T_{\mathrm{sym}}(\mu) = T_{\mathrm{slot}}(\mu)/14$. Semantic adaptation then occupies mini-slot-like grants of $n_{\mathrm{sym},t} \in \{2,4,7\}$ symbols that are dynamically provisioned to the semantic slice. If $\kappa_t$ such grants are allocated and $T_{\mathrm{ctrl},t}$ captures control overhead, the available processing window is
$T_{\mathrm{avail},t} = \kappa_t n_{\mathrm{sym},t} T_{\mathrm{sym}}(\mu) - T_{\mathrm{ctrl},t}.$
Let $b_{i,t} \in \{0,1\}$ indicate whether UE $i$ is scheduled for adaptation in frame $t$. The aggregate time spent by the selected UEs must satisfy
\begin{equation}
\sum_{i=1}^{N} b_{i,t} \, C^{\mathrm{RIC}}_{i,t}(u_t) \le T_{\mathrm{avail},t}.
\end{equation}
Each per-UE near-RT RIC queue integrates arrivals and service
\begin{equation}
Q_{i,t+1} = \big[ Q_{i,t} + A_{i,t} - b_{i,t} C^{\mathrm{RIC}}_{i,t}(u_t) \big]^+,
\end{equation}
where $A_{i,t}$ is an arrival process capturing newly arrived adaptation jobs (e.g., fresh feedback or analytics-triggered updates). Maintaining $Q_{i,t} \le Q_i^{\max}$ mitigates deadline violations and provides additional state information to the CMDP agent.

\section{CMDP-Based Constrained Policy Optimization}
\label{sec:method}

\subsection{CMDP Problem Setup}
\label{subsec:CMDP}
We formulate the time-constrained semantic adaptation problem as a CMDP $\mathcal{M} = (\mathcal{S}, \mathcal{A}, P, r, \mathbf{c}, \gamma)$
whose components reflect the semantics, human feedback, and latency dynamics developed in Sec.~\ref{sec: system model}. 

\boldpar{State space} At the beginning of frame $t$, the near-RT RIC observes the state
\begin{equation}
\begin{aligned}
s_t = \big(&\mathbf{q}_t, \bar{\mathbf{U}}_t, \Delta_t, \boldsymbol{\delta}_t,
\mathbf{Q}_t, \mathbf{H}_t, T_{\mathrm{avail},t}\big),
\end{aligned}
\end{equation}
whose elements are:
\begin{itemize}[leftmargin=*]
    \item $\mathbf{q}_t = [q_{1,t},\ldots,q_{N,t}]^\top$: instantaneous semantic quality seen by the UEs.
    \item $\bar{\mathbf{U}}_t = [\bar{U}_{1,t},\ldots,\bar{U}_{N,t}]^\top$: human-aligned utility estimates maintained by the RIC.
    \item $\Delta_t = [\Delta_{1,t},\ldots,\Delta_{N,t}]^\top$ and $\boldsymbol{\delta}_t = [\delta_{1,t},\ldots,\delta_{N,t}]^\top$: residual slack and normalized deadline debt per UE.
    \item $\mathbf{Q}_t = [Q_{1,t},\ldots,Q_{N,t}]^\top$: RIC queue backlogs of pending adaptation jobs.
    \item $\mathbf{H}_t = \{\mathbf{H}_{i,t}\}_{i=1}^N$: effective channel matrices influencing shared-link reliability.
    \item $T_{\mathrm{avail},t}$: mini-slot–like budget granted to the semantic slice in frame $t$.
\end{itemize}


\boldpar{Action space} 
The agent chooses a composite action
$a_t = (u_t,\mathbf{b}_t)$,
where $u_t \in \mathcal{U} =$ \{\textsc{LightAdapt}, \textsc{FeatRefine}, \textsc{FullRetrain}, \textsc{DeployCached}, \textsc{NoOp}\}
selects the adaptation primitive, and $\mathbf{b}_t = [b_{1,t},\ldots,b_{N,t}]^\top \in \{0,1\}^N$ schedules the UEs that will execute the primitive in frame $t$. In practice, \textsc{LightAdapt} refreshes decoder statistics or adapters with minimal latency, \textsc{FeatRefine} performs moderate-strength fine-tuning (e.g., LoRA layers), \textsc{FullRetrain} deploys a heavy knowledge-base update, and \textsc{DeployCached} rolls back or reinstates a cached stable model, while \textsc{NoOp} leaves the current configuration untouched. The instantaneous feasibility region is
\begin{equation}
\mathcal{A}_{\mathrm{feas}}(s_t)
= \left\{(u,\mathbf{b}) \,\middle|\,
\begin{aligned}
&\sum_{i=1}^{N} b_i C^{\mathrm{RIC}}_{i,t}(u) \le T_{\mathrm{avail},t},\\
&C_{i,t}(u) \le d_i \quad \text{if}\ b_i=1
\end{aligned}
\right\},
\end{equation}
which ensures that the near-RT RIC processing window and every UE deadline remain satisfied.

\boldpar{Transition Kernel}
Given $(s_t,a_t)$, the next state $s_{t+1}$ is sampled from $P(s_{t+1}|s_t,a_t)$. The kernel encompasses: (i) JSCC reconstructions through the fading channels $\mathbf{H}_{i,t}$; (ii) human-feedback fusion producing $\bar{U}_{i,t}$; and (iii) the latency decomposition that updates $\Delta_{i,t}$ and $Q_{i,t}$ via the components of $C_{i,t}(u)$ detailed in Sec.~\ref{sec: system model}. 

\boldpar{Reward and Constraint Signals}
The per-frame reward balances semantic improvement with the computational overhead of large updates:
\begin{equation}
r(s_t,a_t) = \sum_{i=1}^N w_i \bar{U}_{i,t+1} - \beta_u \chi(u_t) - \beta_\delta \sum_{i=1}^N \delta_{i,t+1},
\end{equation}
where $w_i$ encode service priorities, $\chi(u_t)$ quantifies the compute cost at the gNB/RIC, and $\beta_u,\beta_\delta \ge 0$ modulate the trade-off between semantic gain and deadline stress.

We adopt post-transition rewards that depend on $s_{t+1}$ (via $\bar{U}_{i,t+1}$ and $\delta_{i,t+1}$) to couple the learning targets with observed outcomes.
Two cost signals monitor resource feasibility:
\begin{align}
c^{(1)}(s_t,a_t) &= \sum_{i=1}^{N} b_{i,t} C^{\mathrm{RIC}}_{i,t}(u_t),\\
c^{(2)}(s_t,a_t) &= \sum_{i=1}^{N} [C_{i,t}(u_t)-d_i]^+,
\end{align}
representing the near-RT RIC processing time consumed in frame $t$ and the aggregate deadline overshoot.

\boldpar{Optimization Objective}
Let $\pi_\theta(a|s)$ be a stationary stochastic policy with parameters $\theta$ and discount factor $\gamma \in (0,1)$. The CMDP seeks
\begin{subequations}
\label{eq:cmdp_opt}
\begin{align}
\max_{\pi_\theta} \quad & \mathbb{E}_{\pi_\theta} \!\left[ \sum_{t=0}^{\infty} \gamma^{t} r(s_t,a_t) \right] \\
\text{s.t.} \quad & \limsup_{T \to \infty} \frac{1}{T} \mathbb{E}_{\pi_\theta} \!\left[ \sum_{t=0}^{T-1} c^{(j)}(s_t,a_t) \right] \le d^{(j)}, \quad j=1,2,
\end{align}
\end{subequations}
where $d^{(1)} \triangleq \mathbb{E}[T_{\mathrm{avail},t}]$ denotes the long-term RIC processing budget and $d^{(2)} \triangleq 0$ enforces zero average deadline violations (a relaxed $d^{(2)} > 0$ encodes a tolerable violation probability). Solving \eqref{eq:cmdp_opt} produces a policy that maximizes human-aligned semantic utility while respecting both resource and latency constraints.

\subsection{Primal--Dual PPO Surrogate}
We solve \eqref{eq:cmdp_opt} using a primal--dual variant of PPO that maintains the clipped surrogate structure while introducing Lagrange multipliers $\boldsymbol{\lambda}=[\lambda_1,\lambda_2]^\top$ for the constraints. The stochastic Lagrangian is
\begin{equation}
\begin{aligned}
\mathcal{L}(\theta,\boldsymbol{\lambda}) = \mathbb{E}_{\pi_\theta}\!\Bigg[\sum_{t=0}^{\infty} \gamma^t \Big(
& r(s_t,a_t) \\
& - \sum_{j=1}^{2}\lambda_j \big(c^{(j)}(s_t,a_t)-d^{(j)}\big)
\Big)\Bigg].
\end{aligned}
\end{equation}
A clipped mini-batch surrogate for gradient ascent is
\begin{equation}
\begin{aligned}
L_{\mathcal{B}}(\theta,\boldsymbol{\lambda})
= \frac{1}{|\mathcal{B}|} \sum_{t \in \mathcal{B}} \Big[
& \min\big(\rho_t \hat{A}^{r}_t,\mathrm{clip}(\rho_t,1-\epsilon,1+\epsilon)\hat{A}^{r}_t\big) \\
& - \sum_{j=1}^{2}\lambda_j \hat{A}^{c^{(j)}}_t
\Big],
\end{aligned}
\label{eq:ppo_surrogate}
\end{equation}
where $\rho_t=\frac{\pi_\theta(a_t|s_t)}{\pi_{\theta_{\mathrm{old}}}(a_t|s_t)}$ and $\epsilon$ controls the clipping region.

\subsection{Critic Learning and Advantage Estimation}
Low-variance generalized advantage estimates (GAE) are obtained from value baselines.  
The reward critic $V_\psi^r$ minimizes
\begin{equation}
\mathcal{J}_r(\psi)=\tfrac{1}{2|\mathcal{B}|}\sum_{t\in\mathcal{B}}(V_\psi^r(s_t)-\hat{R}_t)^2,
\end{equation}
with targets $\hat{R}_t=\sum_{\ell=0}^{L-1}\gamma^\ell r_{t+\ell}+\gamma^L V_\psi^r(s_{t+L})$.  
Two cost critics $V_{\nu_j}^{c^{(j)}}$ are trained analogously using discounted cost rollouts.  
Advantages use the GAE recursion, e.g.
\begin{equation}
\hat{A}^r_t=\sum_{\ell=0}^{L-1}(\gamma\lambda_{\mathrm{GAE}})^\ell\delta^r_{t+\ell},\quad
\delta^r_t=r_t+\gamma V_\psi^r(s_{t+1})-V_\psi^r(s_t),
\end{equation}
and similarly for $\hat{A}^{c^{(j)}}_t$.  
These baselines embed all state variables so that gradients reflect the coupling between human utility, slack, and queuing delays.

\subsection{Dual Updates and Long-Term Guarantees}
After each policy update, the dual variables ascend along constraint gradients:
\begin{equation}
\lambda_j \!\leftarrow\! [\lambda_j + \eta_\lambda(\hat{c}^{(j)}_{\mathcal{B}} - d^{(j)})]_+, \quad
\hat{c}^{(j)}_{\mathcal{B}} = \tfrac{1}{|\mathcal{B}|}\sum_{t\in\mathcal{B}} c^{(j)}(s_t,a_t).
\label{eq:dual_update}
\end{equation}
Projection onto $\mathbb{R}_+$ enforces non-negativity, and exponential smoothing filters stochastic noise.  
Convergence of the primal–dual iterates implies that the learned policy satisfies the long-term resource and latency budgets.

\begin{figure}[!t]
\centering
\begin{minipage}{0.98\linewidth}
\begin{algorithm}[H]
    \caption{TC-HITL-RL Constrained PPO}
    \label{alg:tchitl}
    \begin{algorithmic}[1]
        \State {Initialize policy parameters $\theta$, reward critic $\psi$, cost critics $\nu_1,\nu_2$, dual multipliers $\lambda_1,\lambda_2$, and empty buffer $\mathcal{D}$.}
        \For{{each training iteration}}
            \State {Collect $L$-frame rollouts using $\Pi_{\mathcal{A}_{\mathrm{feas}}}(\pi_\theta)$ to ensure feasible actions and store transitions in $\mathcal{D}$.}
            \State {Compute advantages $\hat{A}^{r}$, $\hat{A}^{c^{(j)}}$ and returns $\hat{R}$, $\hat{C}^{(j)}$ from the buffered trajectories.}
            \State {Update reward and cost critics via gradient descent on $\mathcal{J}_r$ and $\mathcal{J}_{c^{(j)}}$.}
            \State {Update $\theta$ by maximizing $L_{\mathcal{B}}(\theta,\boldsymbol{\lambda})$ \eqref{eq:ppo_surrogate} with PPO-style clipped gradients.}
            \State {Ascend dual variables: $\lambda_j$ 
            \eqref{eq:dual_update}, followed by exponential moving average (EMA) smoothing.}
            \State {Refresh safety-layer latency predictors using the new latency observations if available.}
        \EndFor
        \State {Deploy the final policy with online shielding for inference.}
    \end{algorithmic}
\end{algorithm}
\end{minipage}
\end{figure}
\subsection{Action Shielding for Instantaneous Feasibility}
\label{subsec:action shield}
It should be noted that average constraints cannot guarantee per-frame safety, so the near-RT RIC implements an action shield that maps tentative outputs of $\pi_\theta$ to a feasible pair in $\mathcal{A}_{\mathrm{feas}}(s_t)$ via a discrete projection: keep the primitive $u_t$ whenever feasible and greedily prune the scheduling mask $\mathbf{b}_t$; if no feasible mask exists, back off to the next lighter primitive, repeating until feasibility or \textsc{NoOp}.
Concretely, a latency-aware knapsack routine prunes the mask based on residual slack or queue urgency; if still infeasible, the shield backs off to a lighter primitive or \textsc{NoOp}.
This ensures per-frame real-time feasibility during both training and deployment.

Algorithm~\ref{alg:tchitl} summarizes the training pipeline, highlighting how rollouts, critic learning, dual updates, and action shielding interact within the TC-HITL-RL framework.

\section{Simulation and Results}

\subsection{Simulation Setup}
We simulate a single semantic-aware gNB serving $N\!\in\!\{8,16\}$ UEs with heterogeneous deadlines and backlog. Each 10\,ms interval samples a numerology $\mu\!\in\!\{0,1,2\}$, grants minislot resources, and exposes the CMDP state variables in Sec.~\ref{sec: system model}; all radio, latency, and learning hyperparameters are fixed across runs and listed in Table~\ref{tab:sim_params}, with configuration files released alongside the code.

The agent selects among the primitives in Sec.~\ref{subsec:CMDP} (\textsc{FullRetrain}, \textsc{FeatRefine}, \textsc{LightAdapt}, \textsc{DeployCached}, \textsc{NoOp}), while the safety shield of Sec.~\ref{subsec:action shield} prunes infeasible UE masks and falls back along that order until per-user timing constraints are met. Latency components follow Sec.~\ref{sec: system model} but inject stochastic congestion (backlog-driven) and fading penalties, and human feedback is drawn from a noisy oracle that rewards semantic quality yet penalises tardy updates, shaping the CMDP reward/cost landscape.


\begin{figure}[!t]
    \centering
    \begin{subfigure}[t]{0.48\linewidth}
        \centering
        \includegraphics[width=\linewidth]{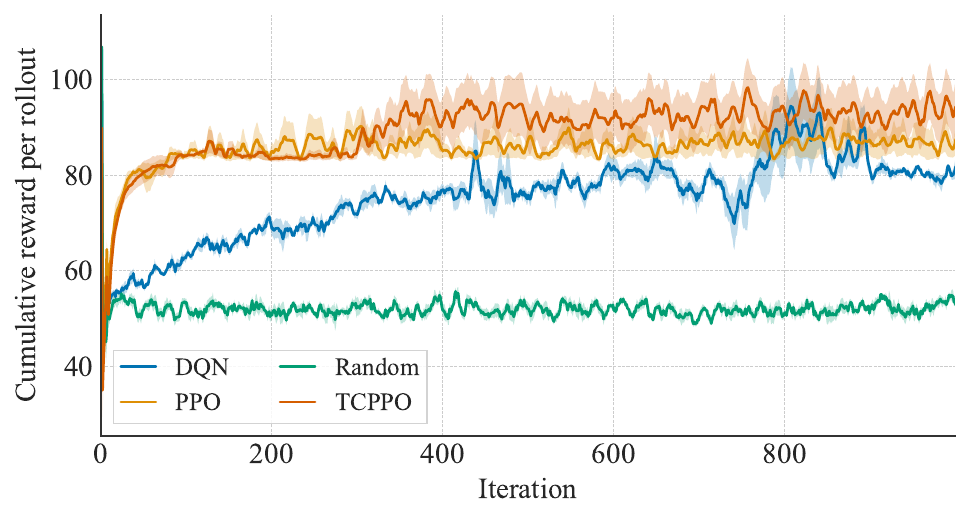}
        \caption{}
        \label{fig:training_reward_U8}
    \end{subfigure}
    \hfill
    \begin{subfigure}[t]{0.48\linewidth}
        \centering
        \includegraphics[width=\linewidth]{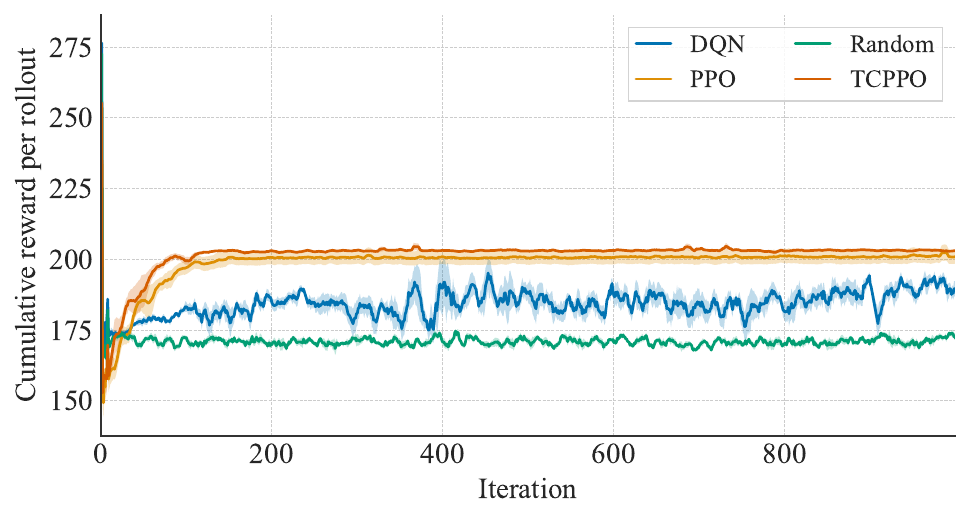}
        \caption{}
        \label{fig:training_reward_U16}
    \end{subfigure}
    \caption{Training reward trajectories (mean $\pm$ std). (a) corresponds to $N\!=\!8$, (b) to $N\!=\!16$.}
    \label{fig:training_reward}
\end{figure}

\begin{figure}[!t]
    \centering
    \begin{subfigure}[t]{0.48\linewidth}
        \centering
        \includegraphics[width=\linewidth]{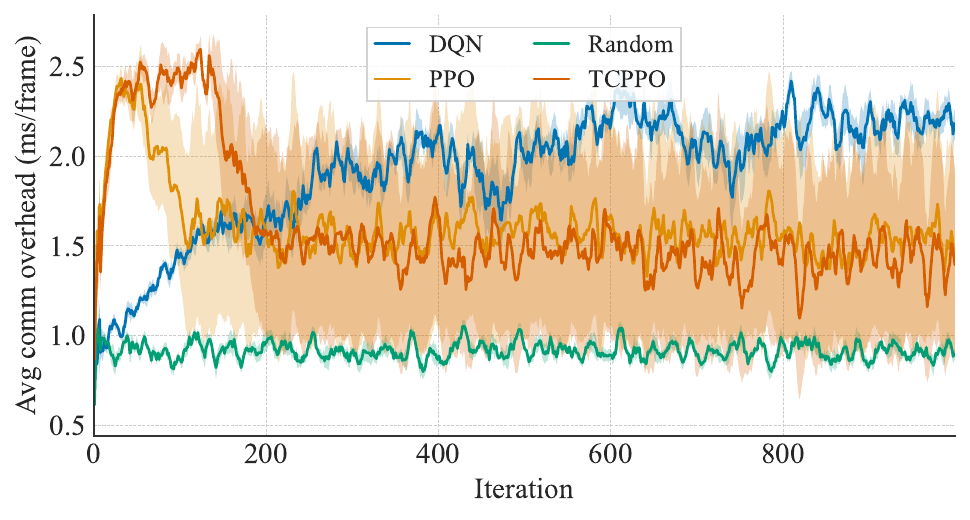}
        \caption{}
        \label{fig:air_U8}
    \end{subfigure}
    \hfill
    \begin{subfigure}[t]{0.48\linewidth}
        \centering
        \includegraphics[width=\linewidth]{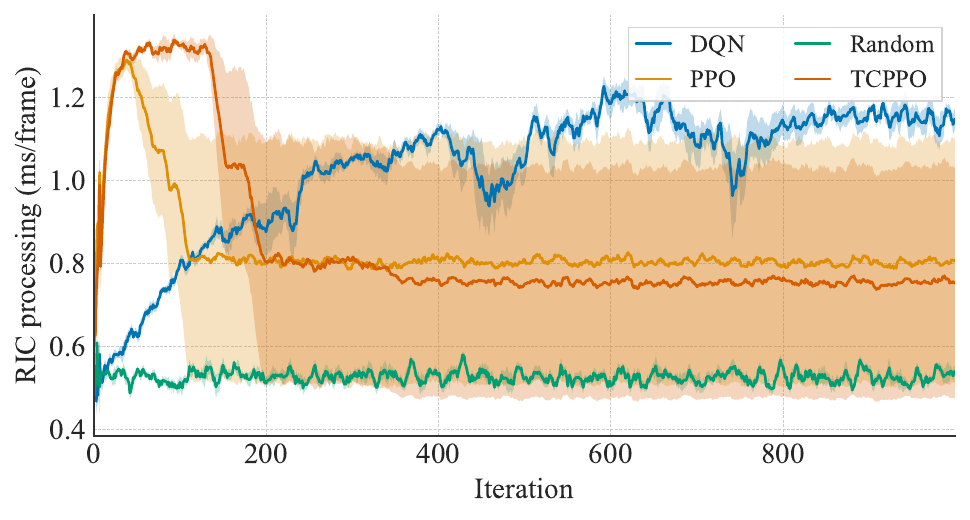}
        \caption{}
        \label{fig:ric_U8}
    \end{subfigure}

    \begin{subfigure}[t]{0.48\linewidth}
        \centering
        \includegraphics[width=\linewidth]{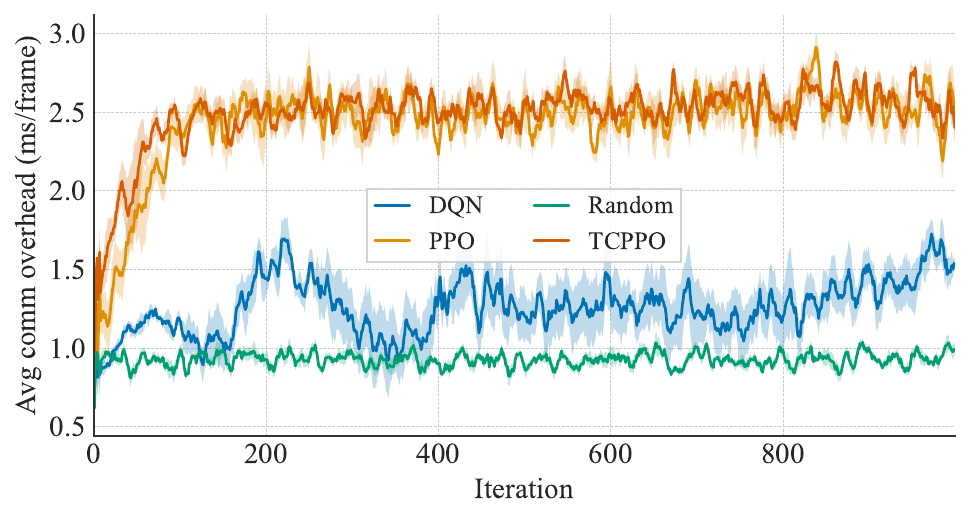}
        \caption{}
        \label{fig:air_U16}
    \end{subfigure}
    \hfill
    \begin{subfigure}[t]{0.48\linewidth}
        \centering
        \includegraphics[width=\linewidth]{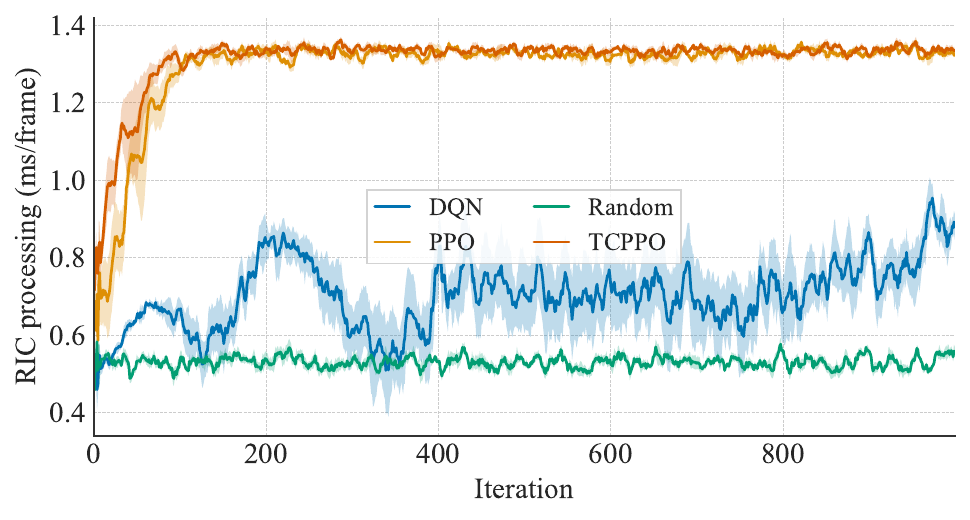}
        \caption{}
        \label{fig:ric_U16}
    \end{subfigure}

    \caption{Average communication and RIC processing budgets consumed during training. (a)--(b) correspond to $N\!=\!8$ and (c)--(d) to $N\!=\!16$.}
    \label{fig:training_overhead}
\end{figure}

For learning, the constrained PPO agent employs two shared-hidden-layer MLPs for the policy and critics; the associated optimization hyperparameters follow the values listed in Table~\ref{tab:sim_params}.

\begin{table}[!t]
    \caption{Simulation parameters}
    \label{tab:sim_params}
    \centering
    \scriptsize
    \begin{tabular}{p{2.1cm}p{5.2cm}}
        \toprule
        Category & Value/Setting \\
        \midrule
        Topology & $N \in \{8,16\}$ UEs with deadlines $d_i \in [6,12]$\,ms. \\
        Radio resources & Scenario-specific numerology $\mu\in\{0,1,2\}$; mini-slot-like grants with $n_{\mathrm{sym}}\in\{2,4,7\}$ symbols; Gaussian $T_{\mathrm{ctrl}}$ for HARQ/CSI overhead. \\
        Primitive latencies (ms) & RIC: $\{5.0, 2.8, 1.1, 1.5, 0\}$; end-to-end: $\{8.4, 5.0, 2.4, 3.1, 0.1\}$ for (\textsc{FullRetrain}, \textsc{FeatRefine}, \textsc{LightAdapt}, \textsc{DeployCached}, \textsc{NoOp}). \\
        Semantic gain means & $\{0.028, 0.019, 0.011, 0.014, 0\}$ for the above primitives; compute penalties $\chi(u)\in\{2.7, 1.6, 0.7, 0.9, 0\}$. \\
        Learning setup & Two shared hidden layers (256/128, ReLU); learning rates $3\!\times\!10^{-4}$; $\epsilon=0.2$; $\lambda_{\mathrm{GAE}}=0.95$; $\gamma=0.99$; dual step size $10^{-3}$ with EMA $0.9$; rollout length $L=64$; mini-batch size $256$; 120 updates. \\
        Seeds & Five seeds per agent ($42$--$46$). \\
        \bottomrule
    \end{tabular}
\end{table}

We benchmark TC-PPO against three baselines implemented in the shared simulation harness: (i) an unconstrained PPO variant with deactivated multipliers, (ii) a discrete-action DQN scheduler whose actions index primitive--mask templates built from slack/backlog statistics, and (iii) a random feasible scheduler \footnote{Unless noted, all results aggregate five seeds per agent (seeds $42$--$46$) and report mean trajectories with standard-error bands.}. To keep this DQN baseline from collapsing into near-idle schedules, we introduce a light under-utilisation penalty that subtracts reward whenever the reported communication or RIC budgets fall below preset service targets, ensuring that every agent is evaluated under comparable minimum service levels.

\subsection{Results}
Fig.~\ref{fig:training_reward} compares the mean training reward for $N\!\in\!\{8,16\}$ users over five seeds. The unconstrained PPO and the constrained TC-PPO converge within approximately $200$ iterations and sustain the highest semantic utility. 
For the DQN baseline, because it lacks dual updates and only selects among pre-defined schedule templates, it plateaus below the PPO variants even though it now activates multi-user masks. Random scheduler remains well below the learning-based methods. The resulting traces emphasise that the DQN regulariser enforces comparable service levels, so the observed reward gap is due to the absence of CMDP guidance rather than artificially low utilisation.

Fig.~\ref{fig:training_overhead} reports the associated air-interface overhead $C^{\mathrm{fb}}\!+\!C^{\mathrm{tx}}$ and the near-RT RIC processing time $C^{\mathrm{RIC}}$. When $N\!=\!8$, PPO/TC-PPO opportunistically alternate between \textsc{FullRetrain} and light-weight primitives whenever slack permits, producing larger error bars but unlocking higher reward. At $N\!=\!16$, the same agents must remain in heavy-update mode and the overhead curves stabilise. DQN, driven by the utilisation penalty, now consumes communication time that is comparable to TC-PPO while still lagging in reward, whereas the Random scheduler remains the only option with consistently low overhead. The right-hand panels further show that TC-PPO tracks PPO’s average RIC consumption but with visibly tighter bands, indicating that enforcing latency constraints also stabilises near-RT compute demand.

\begin{figure*}[!t]
    \centering
    \begin{subfigure}[t]{0.49\linewidth}
        \centering
        \includegraphics[width=\linewidth]{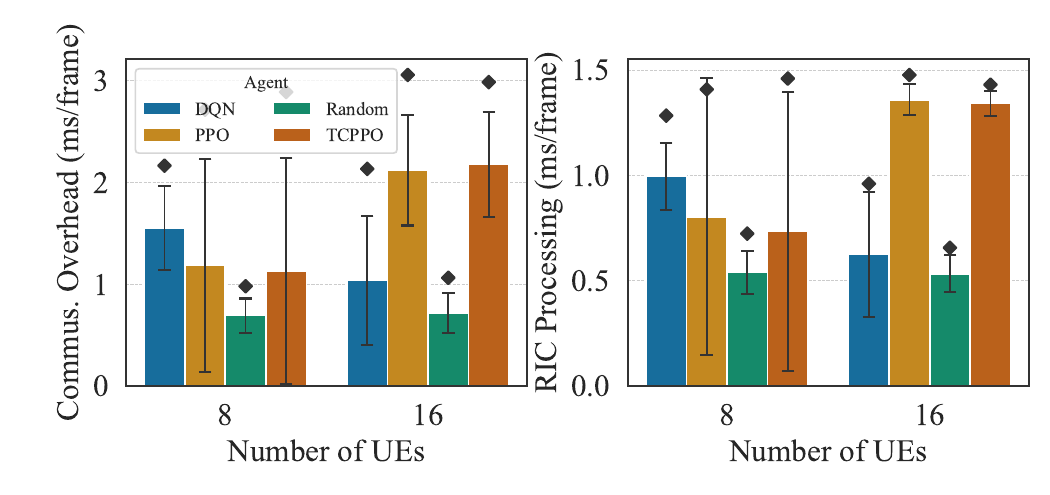}
        \caption{}
        \label{fig:eval_var_resources}
    \end{subfigure}
    \hfill
    \begin{subfigure}[t]{0.49\linewidth}
        \centering
        \includegraphics[width=\linewidth]{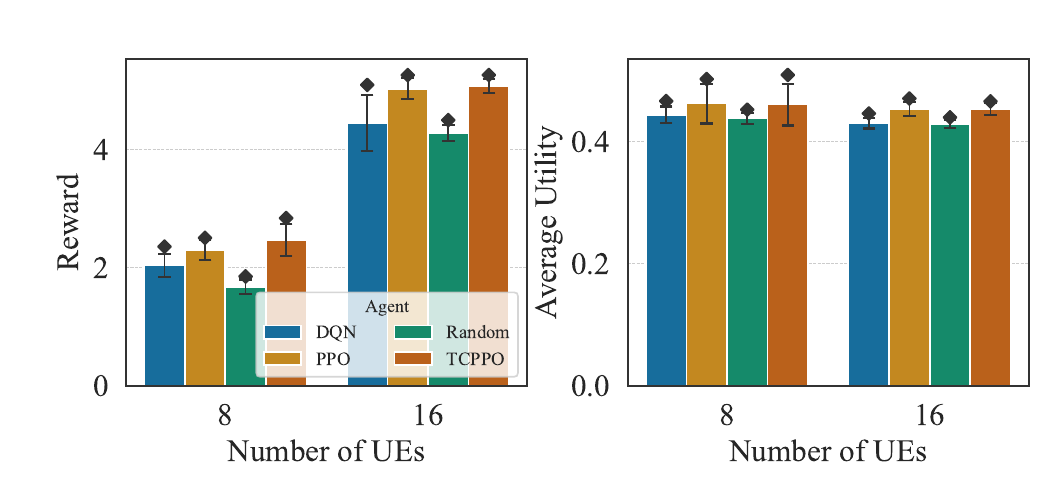}
        \caption{}
        \label{fig:eval_var_reward}
    \end{subfigure}

    \caption{Inference-phase variability across 30 evaluation episodes (mean, SE, and 95th percentile). 
(a) Communication and RIC resource dispersion; (b) reward and utility. TC-PPO matches PPO's reward while stabilizing resource usage.}
    \label{fig:eval_variability}
\end{figure*}

To quantify deployment stability of each agent, Fig.~\ref{fig:eval_variability} plots the metrics aggregated $30$ inference episodes after training and log the per-episode metrics and visualizes their variability. TC-PPO matches the reward of PPO at both user densities while keeping air-interface and RIC overhead within comparable or tighter dispersion bands; all episodes satisfy the deadline constraints (hit rate $=1$). DQN and Random deliver lower reward/utility while the former expends resource budgets similar to TC-PPO because of the minimum-service regulariser.
These results confirm that TC-PPO attains the semantic benefit of PPO with predictable resource usage during deployment. 

\begin{figure}[!t]
    \centering
    \includegraphics[width=0.9\linewidth]{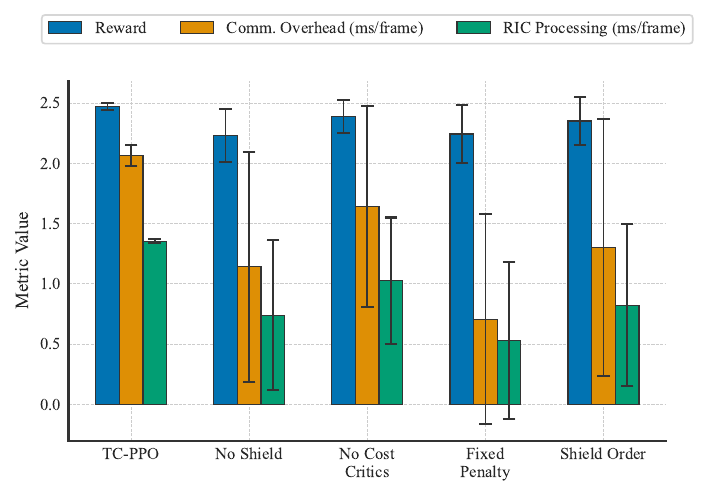}
    \caption{Ablation comparison for $N\!=\!8$ (mean over five seeds): reward, air-interface overhead, and RIC processing.}
    \label{fig:ablation}
\end{figure}

\boldpar{Ablation study}
We further dissect the framework by disabling four key components: (i) removing the safety shield and relying solely on the average CMDP constraints, (ii) removing the cost critics so that timing penalties are injected directly into the reward, (iii) freezing the dual multipliers to a constant penalty, and (iv) reversing the shield fallback order (Light$\!\to$Feat$\!\to$Full). Fig.~\ref{fig:ablation} summarises the evaluation-phase averages (five seeds, $N\!=\!8$). Eliminating the shield notably reduces communication/RIC budgets but yields the lowest reward because the dual controller alone cannot prevent aggressive updates. Penalty-only learning converges but exhibits higher resource variance than the dual formulation, whereas fixed multipliers over-constrain the policy. Reversing the shield order prioritises lightweight updates and trades reward for reduced overhead, underscoring the complementary nature of average constraints and instantaneous feasibility.

\section{Discussion and conclusion}
TC-HITL-RL shows that semantic adaptation can satisfy both average and per-frame latency requirements without sacrificing reward, yet several assumptions merit future work. Reliable and low-latency uplink feedback is presumed; deploying in congested environments will require prioritized bearers, lightweight compression, and deadline-aware buffering. Our evaluation focuses on a single cell, so multi-cell and cooperative-edge scenarios, with coupled RICs and shared fronthaul limits, remain open problems. Finally, we collect scalar human preferences; richer multi-dimensional feedback with confidence scores could further sharpen the CMDP state. Addressing these limitations, along with hardware-in-the-loop validation on NR testbeds, forms our next research agenda.

Within this scope we presented a TC-HITL-RL framework that casts semantic broadcasting as a CMDP and solves it with a primal--dual PPO agent plus an action shield. The resulting policy maintains PPO-level reward while tightening the dispersion of air-interface and RIC resources; ablations confirm the complementary impact of dual updates, shielding, and primitive design. These findings suggest that principled CMDP control is a promising path toward deployable, latency-aware semantic communication.


\section*{Acknowledgment}
This work is supported by the 6G-GOALS project under the 6G SNS-JU Horizon program, n.101139232.
\bibliographystyle{IEEEtran} %
\bibliography{IEEEabrv,references} 
\end{document}